\documentstyle[emulateapj]{article}

\def\drv#1#2{{d #1 \over d #2}}
\def\tdrv2#1#2{{d^2 #1\over d{#2}^2}}

\def\Myr{{\rm\,Myr}}

\def\Gyr{{\rm\,Gyr}}

\def\kms{{\rm\,km\,s^{-1}}}
\def\kpc{{\rm\,kpc}}

\def\pc{{\rm\,pc}}
\def\cm{{\rm\,cm}}
\def\yr{{\rm\,yr}}

\def\spose#1{\hbox to 0pt{#1\hss}}
\def\lta{\mathrel{\spose{\lower 3pt\hbox{$\mathchar"218$}}
     \raise 2.0pt\hbox{$\mathchar"13C$}}}
\def\gta{\mathrel{\spose{\lower 3pt\hbox{$\mathchar"218$}}
     \raise 2.0pt\hbox{$\mathchar"13E$}}}

\input psfig.sty

\def\bfr {{\bf r}}

\newcommand{\half}{{\textstyle{1\over2}}}
\def\K {{\rm\,K}}
\def\eV {{\rm\,eV}}
\def\erg {{\rm\,erg}}
\def\s {{\rm\,s}}

\begin{document}

\title{The Magellanic Stream
	and the density of coronal gas in the Galactic halo}
\author{Chigurupati Murali}
\affil{Canadian Institute for Theoretical Astrophysics \\
University of Toronto, Toronto, ON M5S 3H8, Canada}
\authoremail{murali@cita.utoronto.ca}

\begin{abstract}
The properties of the Magellanic Stream constrain the density of
coronal gas in the distant Galactic halo.  We show that motion through
ambient gas can strongly heat Stream clouds, driving mass loss and
causing evaporation.  If the ambient gas density is too high, then
evaporation occurs on unreasonably short timescales.  Since heating
dominates drag, tidal stripping appears to be responsible for
producing the Stream.  Requiring the survival of the cloud MS IV for
$500\Myr$ sets an upper limit on the halo gas density $n_H<
10^{-5}\cm^{-3}$ at $50\kpc$, roughly a factor of 10 lower than that
estimated from the drag model of Moore \& Davis (1994).  Implications
for models of the evolution of gas in galaxy halos are discussed.
\end{abstract}


\section{Introduction}
In current pictures of hierarchical galaxy formation, the initial
collapse and continuing accretion of gas-rich fragments produces and
maintains an extended halo of diffuse, hot gas surrounding the galaxy
(White \& Rees 1978; White \& Frenk 1991).  This gaseous halo fills
the dark matter potential, and is roughly in hydrostatic equilibrium
out to distances of order the virial radius.  The inner, more dense
regions cool through thermal brehmsstrahlung and slowly accrete into
the central regions of galaxies.  In the Milky Way, this scenario
predicts gas at a temperature $T_H\sim 10^6\K$ at distances $R\gta
50\kpc$ from the Galactic center.

It is difficult to identify this gas directly from X-ray observations
because of the difficulty in determining distances (Snowden 1998).
Consequently, although it is clear that there is a diffuse background
in the $0.1-2.0$-keV range, it is very difficult to determine how much
of the emission is local (within a few hundred parsecs), from an
extended halo or extragalactic in origin (Snowden 1998; Snowden et
al. 1998).  Although most of the emission at $1/4$-keV does arise
locally (Snowden et al. 1998), most at $3/4$-keV does not.  This
component presumably contains emission originating in the Galactic
halo and in extragalactic sources-- however, the relative
contributions are poorly constrained.  Searches for extensive X-ray
halos around local, late-type galaxies have also proved unsuccesful.
Benson et al. (1999) examined archival ROSAT images of 3 nearby,
massive spirals but detected no emission and established upper limits
which are far below the predicted luminosities (White \& Frenk 1991;
Kauffman et al. 1993; Cole et al. 1994).

Given the difficulty of observing this gas directly, it it useful to
infer its existence and properties indirectly.  Some of the best
evidence comes from the metal absorption lines associated with
galaxies seen in quasar spectra (Bahcall \& Spitzer 1999; Steidel
1998) and also seen in high-velocity clouds in the Milky Way (Sembach
et al. 1999).  The Magellanic Stream offers another potential probe of
hot gas in the Milky Way halo.  The Stream is a long HI filament,
apparently trailing the Magellanic Clouds (Jones, Klemola \& Lin 1994)
and mostly confined to discrete clouds, which are very similar in
properties to other high velocity clouds (Wakker \& van Woerden 1997;
Blitz et al. 1999).  Because it will interact with ambient halo gas,
its observable characteristics should constrain the properties of the
diffuse medium.

Early on, Mathewson et al. (1977) proposed that the Magellanic Stream
represents the turbulent wake of the Magellanic Clouds as they pass
through a diffuse halo medium; however, Bregman (1979) identified a
variety of observational and theoretical difficulties with this model
and concluded that the tidal stripping model (Murai \& Fujimoto 1980;
Lin \& Lynden-Bell 1982; Gardiner \& Noguchi 1996, most recently)
provides a better explanation.  Moore \& Davis (1994) modified the gas
dynamic model to include stripping by an extended ionized disk and
drag by a diffuse halo: their model matches the Stream kinematics well
by incorporating drag from a diffuse gas distribution at $50\kpc$ that
satisfies all known limits.  However, the model remains controversial
(e.g., Wakker \& van Woerden 1997), so that inferences about halo gas
properties are correspondingly uncertain.

In the present paper, we reconcile these competing views of Magellanic
Stream formation and, in doing so, establish limits on the density of
diffuse gas at the current distance of the Magellanic Clouds.  In
brief, we show that the motion of individual Stream clouds through
ambient, ionized gas is dominated not by drag, but by strong heating
from accretion.  The accretion of ambient gas heats the cloud through
thermalization of the bulk flow and through the ionic and electronic
enthalpy of accreted gas.  Weak radiative cooling leads to mass loss
and cloud evaporation.  Requiring cloud survival indicates that only
the tidal stripping model for the Magellanic Stream is viable.
Furthermore, the survival requirement places strong limits on the
density of ionized gas in the halo.  The constraint is roughly an
order-of-magnitude lower than previously inferred.  Discussion of the
cloud-gas interaction and the survival constraint is presented in
\S\ref{sec:limits}.  The implications are examined in \S\ref{sec:imp}.

\section{Limits on halo gas}
\label{sec:limits}
The Magellanic Stream is concentrated primarily in a bead-like
sequence of 6 discrete clouds at high Galactic latitude (Mathewson et
al. 1977).  The cloud MS IV is located near the `tip' of the stream at
$\ell=80^o, b=-70^o$, roughly $60^o$ across the sky from the
Magellanic Clouds (Cohen 1982).  The mean HI column density
$N_H=6\times 10^{19}\cm^{-2}$ (Mathewson et al.  1977), which is
intermediate between the denser clouds that lie close to the LMC and
the more diffuse clouds at the very tip of the Stream.  However, it is
also rather centrally condensed with a peak column density of roughly
$1.3\times 10^{20}\cm^{-2}$ (Cohen 1982).  The cloud has approximate
HI mass $M_c=4500 (d/\kpc)^2 M_{\odot}$, radius $R_c=15 (d/\kpc) \pc$
and temperature $T_c=10^4\K$ as determined from the linewidths (Cohen
1982).  Assuming a pure hydrogen cloud, the mass and radius give a
mean number density $n_c=0.27 (50\kpc/d) \cm^{-3}$.

The kinematics and age of MS IV depend on the formation model.  In the
most recent tidal model (Gardiner \& Noguchi 1996), the eccentrity of
the Magellanic Clouds is relatively small, so that MS IV, having been
tidally stripped at perigalacticon roughly $1.5 \Gyr$ ago and
following nearly the same orbit, would have a velocity of $220\kms$ at
roughly $50\kpc$.  In the gas drag model, Moore \& Davis (1994) argue
that the Stream was torn from the Magellanic Clouds during passage at
$65\kpc$ through an extended, ionized portion of the Galactic disk
roughly $500\Myr$ ago.  From a momentum-conservation argument, they
deduce an initial velocity of $220\kms$ after separation for MS IV.
Additional drag from the ambient halo medium modifies the orbit to
give the current radial velocity of $-140\kms$ with respect to the
local standard of rest at a distance $d\sim 20\kpc$.  The transverse
velocity is unspecified but must be large ($\sim 340\kms$) because
even a total velocity of $\sim 300\kms$ implies that the orbital
energy has diminished by $10^{54}\erg$ since separation.  The
dissipated energy heats the cloud, which has thermal energy $E_c=3/2
M_c/m_h kT_c\approx 10^{51}\erg$ at $20\kpc$-- roughly 0.1\% of the
input energy: the cloud must therefore evaporate.  However, as the
analysis below shows, if we adopt lower bounds on the velocity and age
of MS IV, $V_c=220\kms$ and $t=500\Myr$, we obtain an upper limit on
the gas density at $50\kpc$ which is lower than that required to give
the drag in the Moore \& Davis (1994) model.

In addition to the short evaporation timescale, it is difficult to
accept the smaller distance to MS IV because, at $50\kpc$, the cloud
temperature, mass and size put it approximately in virial equilibrium,
which naturally explains its centrally condensed appearance.  At
$20\kpc$, the cloud should be unbound and rapidly expanding, unless
confined by a strong external pressure.

The parameters of the halo gas at either distance are rather
uncertain.  Following current galaxy formation models (e.g., White \&
Frenk 1991), we assume that the gas is in quasi-hydrostatic
equilibrium in the Galactic potential.  The estimated temperature of
the halo $T_H=2.9\times 10^6\K$ for an isothermal halo with rotation
speed $V_0=220\kms$.  This implies that the sound speed $c_H=200\kms$.
Halo gas at this distance may rotate with velocities on the order of
$10-20 \kms$ leading to a small reduction in its temperature or
density.  The rotation would have little influence on the Stream-gas
interaction since gas rotation would be aligned with the disk while
the Stream travels on a nearly polar orbit.

Previously, the density of halo gas has been estimated by Moore \&
Davis (1994) using a drag model to account for the kinematics of the
Stream clouds.  Matching the kinematics of the Stream requires a gas
particle density $n_H\sim 10^{-4}\cm^{-3}$ at a distance of
approximately $50\kpc$.  As noted above, this approach neglects the
energy input into the cloud as it snowplows though the halo
medium\footnote{Although clouds have been modeled as blunt objects
(e.g., Benjamin \& Danly 1997), the boundary conditions are different.
The no-slip and no-penetration boundary conditions are not applicable
since the cloud is permeable.  Magnetic fields do not prevent
penetration and shear stress because of charge transfer.}.  As we will
show below, heating dominates drag and cloud survival against
evaporation sets a much stronger limit on the halo gas density.  This
argument is similar to that posed by Cowie \& McKee (1976) in
constraining the density of ionized gas in the intergalactic medium
based on the timescale for conductive evaporation of neutral clouds.

\subsection{Energy input and mass loss}
In its rest frame, the total instantaneous internal energy of the
cloud $E_c=T+W$, where $T$ is the total thermal energy and $W$ is the
potential energy.  The rate of change in energy is determined by
energy input and loss:
\begin{equation}
\drv{E_c}{t}=\rho_H v_H A (\half v_H^2+{5\over 2} c_H^2)
	-\Lambda n_H n_c V-{\dot M}(\half u^2+{5\over 2}\Delta c^2)+{\dot W}.
\label{eq:edot}
\end{equation}
The first term on the right-hand side gives the energy input through
the projected surface area $A=\pi R_c^2$ of the cloud's leading edge
from halo gas with density $\rho_H$, streaming velocity $v_H$ and
sound speed $c_H$; the second term gives the inelastic cooling rate
with reaction rate $\Lambda$ in the volume $V$ at the cloud surface
where halo gas at density $n_H$ mixes with cloud material at mean
density $n_c$; the third term gives the cooling from cloud mass loss
at surface velocity $u$ and change in enthalpy $5\Delta c^2/2$, where
$\Delta c^2$ is the change in the square of the cloud sound speed; the
last term gives the rate of change of the potential
energy\footnote{Since $W=\int d\bfr \rho\Phi$, $\dot W=\int d\bfr
[\dot\rho\Phi+\rho\dot\Phi]$.  Generally, when the cloud loses mass,
$\dot\rho<0$ and $\dot\Phi>0$ while $\rho>0$ and $\Phi<0$, so that
$\dot W>0$.}

In a steady state, $\dot E_c\approx 0$ (e.g., Cowie \& McKee 1977).
As we discuss below, the protons carry roughly 2/3 of the incident
energy: all of the energy of bulk flow and half of the enthalpy, which
is of the same order.  However, at these energies ($\sim 100\eV$),
proton collisions with cloud HI are dominated by charge transfer:
inelastic losses are negligible\footnote{Charge transfer at these
relative velocities redistributes particle momentum and energy, rather
than creating photons (Janev et al. 1987).}.  Therefore, neglecting
$\dot W$ heating, we obtain the steady-state mass loss rate
\begin{equation}
\dot M={\rho_H v_H A (\half v_H^2+{5\over2} c_H^2)\over
\half u^2+{5\over 2}\Delta c^2}.
\label{eq:mdot}
\end{equation}
The mass loss rate is equal to the accretion rate times the ratio of
specific energy input to specific energy outflow.  The outflow
velocity $u\sim v_e$, the surface escape velocity of the cloud in the
tidal field.  A typical cloud is loosely bound so that $v_e\sim
v_{therm}$, the thermal velocity of the cloud.  The change in enthalpy
is of the same order.  For MS IV, this implies that the evaporated
mass leaves the cloud with $u\sim 10\kms$ and $T\sim 2\times 10^4\K$.
Figure 1 shows the mass loss rate for various combinations of relative
velocity and ambient gas density.  For the minimum relative velocity
$v_H=220\kms$ and age $t=500\Myr$ defined above, $n_H< 10^{-5}\cm^{-3}$
in order for the cloud to survive.

\psfig{file=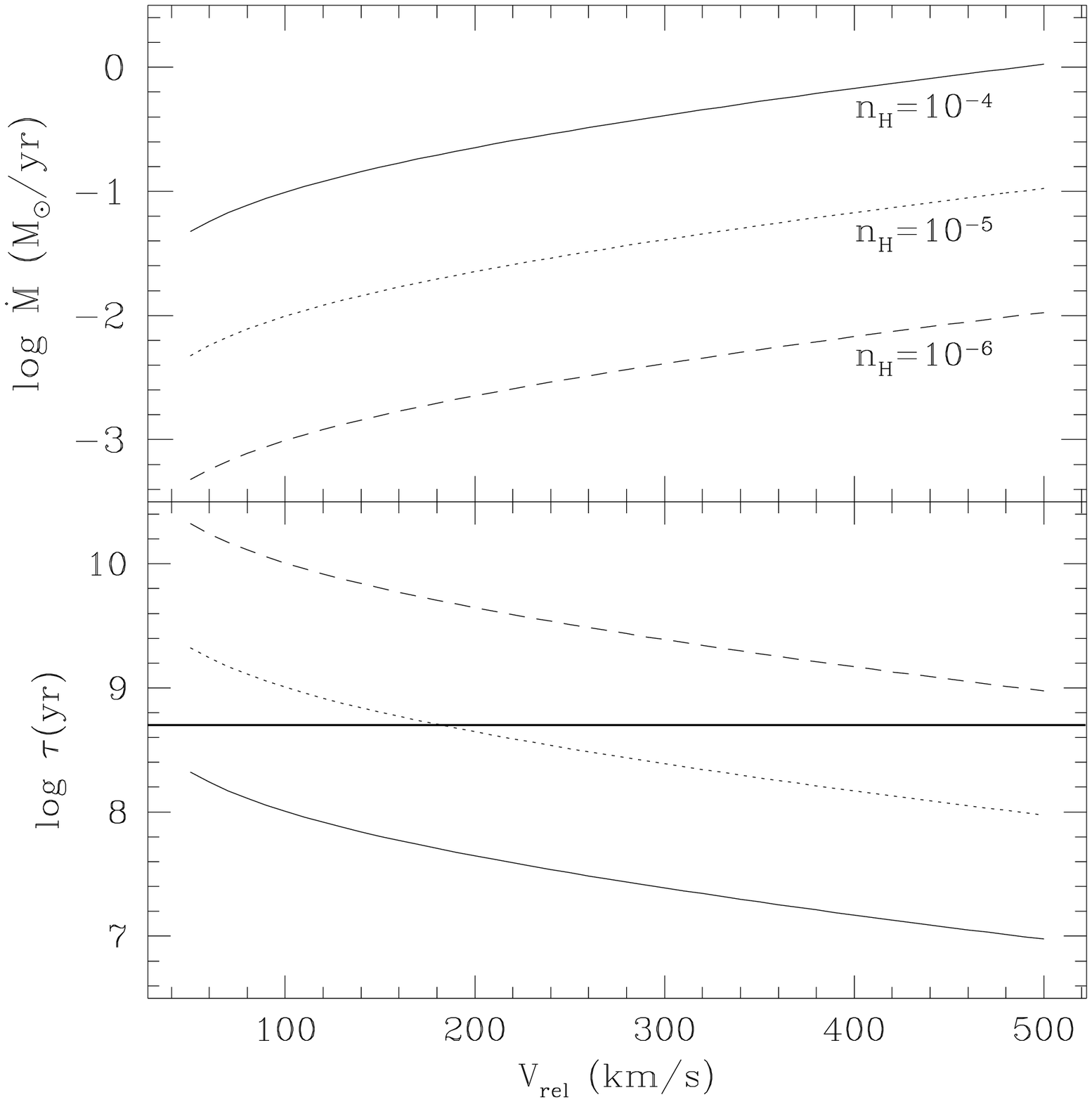,width=9cm}
\smallskip\noindent {\sl Fig.~1~}\ Cloud mass loss rates (upper panel)
and corresponding lifetimes (lower panel) as a function of relative
velocity for different halo densities.  The horizontal line in the
lower panel shows the $500\Myr$ cutoff.  Lifetimes shorter than this
are unlikely.

\bigskip

Mass loss is not spherically symmetric as in the evaporation of cold
clouds embedded in a hot, diffuse medium (Cowie \& Mckee 1977; Balbus
\& Mckee 1982; Draine \& Giuliani 1984).  Here the halo gas strongly
heats the cloud at the leading edge causing outflow along this surface
and ablation of material from the poles (Murali 1999); the interaction
is analogous to that of comets with the solar wind, also referred to
as a {\it mass loaded flow} (Biermann et al. 1967; Wallis \& Ong 1975;
Galeev et al. 1985).  Since the flow of halo gas is roughly
supersonic, a bow shock may form; however, the shock will be very weak
and approximately at or interior to the leading edge of the cloud
because of cooling from mass loading (Wallis \& Ong 1975).
Nevertheless, the morphology may be detectable through sensitive X-ray
or EUV observations.

\subsection{Accretion and drag}
Accretion of mass and momentum has only a small effect on the cloud
even though energy accretion is significant.  The mass accretion rate
\begin{equation}
\dot M_{acc}=\rho_H v_H A,
\end{equation}
while the momentum accretion rate
\begin{equation}
\dot P_{acc}=\rho_H v_H^2 A.
\end{equation}
Note that momentum transfer from accretion is equivalent to drag with
drag coefficent $C_D=2$.  At $v_H=220\kms$ and $n_H=10^{-4}\cm^{-3}$,
the mass accretion rate $\dot M=9.0\times 10^{-4} M_{\odot}\yr^{-1}$.
For an accretion time of $5\times 10^8 \yr$ and neglecting mass loss,
the cloud accretes $\sim 5\times 10^5 M_{\odot}$-- roughly 5\% of its
initial mass.  Momentum transfer through accretion reduces the
velocity by roughly $10\kms$.  For $n_H < 10^{-5}\cm^{-3}$, changes in
mass and momentum are entirely negligible.

\subsection{Thermalization and cooling}
\label{sec:additional}
Collisions between incident protons and electrons in the inflowing
halo gas and target HI in the cloud thermalize the flow and lead to
some excitation and radiative cooling.  Estimates of the relevant
rates can be obtained by considering the cross-sections or reaction
rates for collisions between the incident and target particles given
their densities and typical relative velocity.  Because the halo gas
is so diffuse, $H^+-e^-$ scattering is unimportant: $H-H^+$ and
$H-e^-$ collisions dominate.

Momentum transfer through charge exchange between protons and neutral
hydrogen atoms thermalizes the halo gas flow at the leading edge of
the cloud.  Recent plasma calculations by Krsti\'c \& Schultz (1998)
give momentum transfer cross-sections $\sigma_{mt}$ at the appropriate
energies using the standard method of partial wave expansions to
determine scattering amplitudes (e.g., Landau \& Lifschitz 1977).  For
relative velocities $v_H\sim 200\kms$ or relative energies $\sim
100\eV$, $\sigma_{mt}\sim 10^{-15}-10^{-16} \cm^{-2}$.  Thus the mean
free path into the cloud $\lambda=1/n_c\sigma_{mt}\sim 10^{16}\cm$ for
$n_c\sim 0.25 \cm^{-3}$.

Cooling does little to balance the energy input into the cloud.  Janev
et al. (1987) provides a compendium of thermal reaction rates $\langle
\sigma v_{rel}\rangle$ as a function of relative energy for
excitation, ionization and recombination (inelastic processes) in a
wide range of atomic, electronic and ionic collisions.  Examining
these rates shows that radiation arises purely from collisions between
electrons in the inflowing halo gas and neutral hydrogen atoms in the
cloud.  For $H-H^+$ collisions in a hydrogen plasma at $T=10^4\K=1\eV$
with $v_H=200\kms$, the reaction rate for any inelastic excitation
from ground-state is less than $3\times 10^{-11} \cm^3\s^{-1}$
(Janev et al. 1987, pp 115-136), which is entirely negligible.  Thus,
since the halo gas is so diffuse and energy transfer between electrons
and protons is minimal, all the proton energy (roughly 2/3 of the
total) in the bulk flow heats the cloud; only electron-neutral
collisions can produce radiation.

For ground-state excitation and ionization by electrons under these
conditions, reaction rates are below $5\times 10^{-8}\cm^3\s^{-1}$
(Janev pp. 18-31).  Recombination rates are considerably lower:
$\langle \sigma v_{rel}\rangle <10^{-13} \cm^3 \s^{-1}$ (Janev
pp. 32-33) and cannot be important even after thermalization to the
outflow temperature of $1\eV$.  Thus for densities of incident
particles $n_H\sim 10^{-4}\cm^{-3}$ and target particles $n_c\sim
0.25\cm^{-3}$ with mean relative velocity of order the electron
thermal velocity $v_e$, which is given by the mean energy $100\eV$,
the energy lost to inelastic processes
\begin{equation}
\Lambda n_H n_c V=n_H N_c\sum_i \langle \sigma v_{rel}\rangle_i 
	{\rm h}\bar\nu A < 10^{36} \erg\s^{-1},
\label{eq:emission}
\end{equation}
where $V=A\lambda$ and $N_c=n_c\lambda$, the neutral column density in
the mixing layer.  The sum over $i$ includes the dominant processes:
excitation from $n=1\to n=2(s,p)$, from $n=1\to n=3$ and ionization
from $n=1$ (Janev et al. 1987, pp. 18-27) at a temperature of
$100\eV$.  We take ${\rm h}\bar\nu=10\eV$ and ignore electron cooling
(which reduces the amount of radiation produced) over the mean-free
path of the protons in the cloud; therefore the loss rate is an upper
limit.  After cooling, electrons drop to the thermal velocity of the
outflowing gas; the combination of velocity and density are too low to
permit any additional inelastic cooling.  While the radiation rate is
substantial, it is considerably lower than the total rate of energy
input into the cloud, which is of order $10^{38}\erg\s^{-1}$.
Although Weiner \& Williams (1996) propose that H$\alpha$ from the
leading edge of MS IV can be produced collisionally when $n_H\sim
10^{-4}\cm^{-3}$, the rate estimated here is considerably lower than
measured.  This in turn suggests that escaping UV photons from the
Galactic disk produce the H$\alpha$ emission through ionization and
recombination (Bland-Hawthorn \& Maloney 1999) or possibly through
fluorescence.

\section{Evolution of halo gas}
\label{sec:imp}
In current scenarios of galaxy formation, gas in galactic halos should
cool interior to some radius $r_c$ which increases with time (e.g.,
White \& Rees 1978; Mo \& Miralda-Escud\'e 1996).  For circular
velocity $V_0=220\kms$, $r_c\approx 250\kpc$ at the current
time. However, given the lack of evidence for cooling flows, it is
expected that gas within $r_c$ is heated into a constant entropy core
(Mo \& Miralda-Escud\'e 1996; Pen 1999), so that, for $r<r_c$, the
halo gas density
\begin{equation}
\rho_H(r)=\rho_H(r_c)\biggl[1-{4\over 5}\ln (r/r_c)\biggr]^{3/2}.
\label{eq:rhoh}
\end{equation}
where $\rho_H(r_c)=f_gV_0^2/4\pi G r_c^2$, and $f_g$ is the fraction
of the total halo mass density in gas.  If $f_g$ equals the Universal
baryon fraction, then $f_g\sim 0.05$ for $\Omega_m=1$ and $f_g\sim
0.15$ for $\Omega_m=0.3$, where $\Omega_m$ denotes the ratio of total
mass density to closure density of the Universe.  The constraint on
the density derived here suggests that $f_g\lta 5\times 10^{-3}$.
Within the context of this density model, this discrepancy with the
Universal fraction leads to the possibility that a considerable amount
of gas has cooled and formed stars or dark matter (Mo \&
Miralda-Escud\'e 1996) or has been expelled by strong heating
(e.g., Field \& Perronod 1977; Pen 1999).  Ultimately, however, it is
not clear that this model properly describes the gas distribution in
galactic halos.

\section{Summary}
We have re-examined the interaction of the Magellanic Stream with
ambient gas at large distances in the Galactic halo.  Our analysis
shows that heating dominates over drag.  Therefore, because of their
high relative velocities, clouds are prone to evaporation if the
ambient gas density is too large.  In particular, the requirement that
MS IV survives for $500\Myr$ at $220\kms$ imposes the limit on the
density of halo gas at $50\kpc$: $n_H < 10^{-5} \cm^{-3}$.  This upper
limit is roughly an order-of-magnitude lower than the density
determined from the drag model of Moore \& Davis (1994), and does not
concur with current models of the gas distribution in galactic halos.

\begin{acknowledgments}
I am grateful to Mineo Kimura, Hiro Tawara, Predrag Krsti\'c, Neal
Katz, Gary Ferland and Ira Wasserman for discussion, numerous helpful
suggestions and providing data.  I am also indebted to the referee for
very constructive criticism.  This work was supported by NSERC.
\end{acknowledgments}

\end{document}